\documentclass[11pt,trackchanges,preprint]{aastex63}
\usepackage{natbib}
\usepackage{amssymb,amsbsy,amsmath}
\usepackage{verbatim}
\usepackage{times}
\usepackage{placeins}
\usepackage{aas_macros}
\usepackage[nottoc,numbib]{tocbibind}

\makeatletter
\def\set@curr@file#1{%
  \begingroup
    \escapechar\m@ne
    \xdef\@curr@file{\expandafter\string\csname #1\endcsname}%
  \endgroup
}
\def\quote@name#1{"\quote@@name#1\@gobble""}
\def\quote@@name#1"{#1\quote@@name}
\def\unquote@name#1{\quote@@name#1\@gobble"}
\makeatother

\usepackage{graphicx}

\received{\_\_\_\_\_\_\_\_\_\_}
\revised{\_\_\_\_\_\_\_\_\_\_}
\accepted{\_\_\_\_\_\_\_\_\_\_}

\submitjournal{ApJ}

\shorttitle{3D Reconstruction}
\shortauthors{Plowman}

\begin{document}

\title{Three-dimensional Reconstruction of Coronal Plasma Properties from a Single Perspective}
\correspondingauthor{Joseph Plowman}
\email{jplowman@boulder.swri.edu}

\author[0000-0001-7016-7226]{Joseph Plowman}
\affil{Southwest Research Institute,
Boulder, CO 80302 USA}


	\begin{abstract}
		Much of our understanding of the state of coronal plasmas comes from observations that are optically thin. This means that light travels freely through the corona without being materially affected by it, which allows it to be easily seen through, but also results in a line-of-sight degeneracy which has previously thwarted attempts to recover the three-dimensional structure of the coronal plasma. However, although the corona is disorganized in the line-of-sight direction, it is highly organized in the field-aligned direction. This paper demonstrates how to exploit this organization to resolve the line-of-sight degeneracy in the plasma properties using a suitable magnetic field structure. A preliminary investigation with a potential field is shown, finding a solution which clearly resembles the real solar data, even with a single perspective. The results indicate that there is ample information in the resulting residuals that can be used to refine the magnetic field structure, allowing, for the first time, the optically thin plasma observations to speak directly to the magnetic field extrapolations.
	\end{abstract}


	\section{Introduction}\label{sec:intro}
		
	Radiation from the solar corona in many parts of the electromagnetic spectrum, such as from the Atmospheric Imaging Assembly \citep[AIA;][]{Lemen2012} onboard the \textit{Solar Dynamics Observatory} \citep[SDO;][]{Pesnell2012} or the {\em Hinode} X-ray Telescope \citep[XRT;][]{Golub2008}, is optically thin: Light produced at one point in the corona can pass through the rest of the corona without being meaningfully deflected, absorbed, or otherwise impeded. In particular, the observed coronal radiation along a given line of sight is proportional to the integral of the radiation at each point along that line of sight. This is both a blessing and a curse -- a blessing, because all structures in the corona are equally visible regardless of the coronal material in front of them (with a few exceptions such as prominences); a curse, because it is impossible to directly ascertain any depth information about these structures. Attempts to use multiple perspectives to disentangle this depth information have met with limited success because of the complexity of these structures, limited number of perspectives \citep[e.g., in the case of the STEREO spacecraft;][]{STEREO_Kaiser2008}, and the rapid variation of coronal structures (which impedes attempts to use the solar rotation to provide additional perspectives).
	
	The possibility of recovering the three-dimensional structure, then, might seem to be a pipe dream. And indeed, the problem is formally ill-posed. For every line of sight, a given coronal image provides only a single piece of information, one pixel. Although there are a wide variety of passbands, providing distinct pieces of information for each line of sight, the fact remains all of these pieces of information are completely insensitive to arbitrary rearrangement (or `scrambling') of the volume elements along any given line of sight.
	
	However, the corona is also organized by its magnetic field: in most locations, plasma is forced to spiral around magnetic field lines and cannot move any significant distance perpendicular to it \citep{Reale_LRSP2014}. This provides structure to the plasma which can help to resolve the line-of-sight degeneracy. The magnetic field structure in turn is governed by Maxwell's equations, which enforce constraints on the geometry of the field -- e.g., that its divergence is zero, and the force-free condition, which is approximately true in much of the corona \citep[see][]{WiegelmannSakurai_LRSP2012, MackayYeates_LRSP2012}. These constraints are such that, in principal, the field in a three-dimensional volume can be determined only with measurements at its boundary. Such magnetic field extrapolations have been limited by sensitivity to small variations in the boundary measurements, combined with measurement uncertainty and the difficulty of making vector magnetic field measurements. The field extrapolations, therefore, would greatly benefit from some additional measurements made within the volume. But that is exactly the sort of information that the optically thin measurements can provide.
	
	What is needed, therefore, is to fully bring in the optically thin plasma observations to the problem. Rather than relying solely on noise-sensitive magnetic field extrapolations and/or computationally prohibitive simulations to constrain the physics in the volume, these optically thin data (which respond directly to the plasma state in the volume) should also be used to constrain the volume plasma state. The necessary components are 
	\begin{enumerate}
		\item A means of applying the geometrical constraints that the magnetic field can provide to the optically thin coronal reconstruction problem; {\em and} \label{enum:thinrecon}
		\item A way of applying the volume-sampling information provided by the optically thin measurements to the magnetic field extrapolation problem. \label{enum:extrapconstrain} 
	\end{enumerate}
	This paper describes a newly developed `three dimensional reconstruction' method which solves component \ref{enum:thinrecon}, demonstrates it, and briefly describes how it can be used to solve component \ref{enum:extrapconstrain}. This method allows, for the first time, the full three-dimensional plasma distribution (both temperature and density) to be estimated for an active region or indeed the entire corona.
	

	\section{Optically thin plasma reconstruction core and formalism}\label{sec:formalism}

	At its core, the new three-dimensional reconstruction method uses a magnetic field extrapolation to build a set of linear mappings from the emission of loops (field lines) to the emission observed in pixels of an image or images (e.g., from SDO/AIA). In its most basic form, this can be expressed as a matrix $P_{ij}$: given a vector of coefficients, $c_j$, which scale the overall emission or `intensity' of each loop, the pixel (data) intensities/emissivities ($E_i$) are

	\begin{equation}\label{eq:linear_loop_response}
		E_{i} = \sum_j P_{ij} c_j,
	\end{equation}

\begin{figure}[!htbp]
		\begin{center}\includegraphics[width=0.6\textwidth]{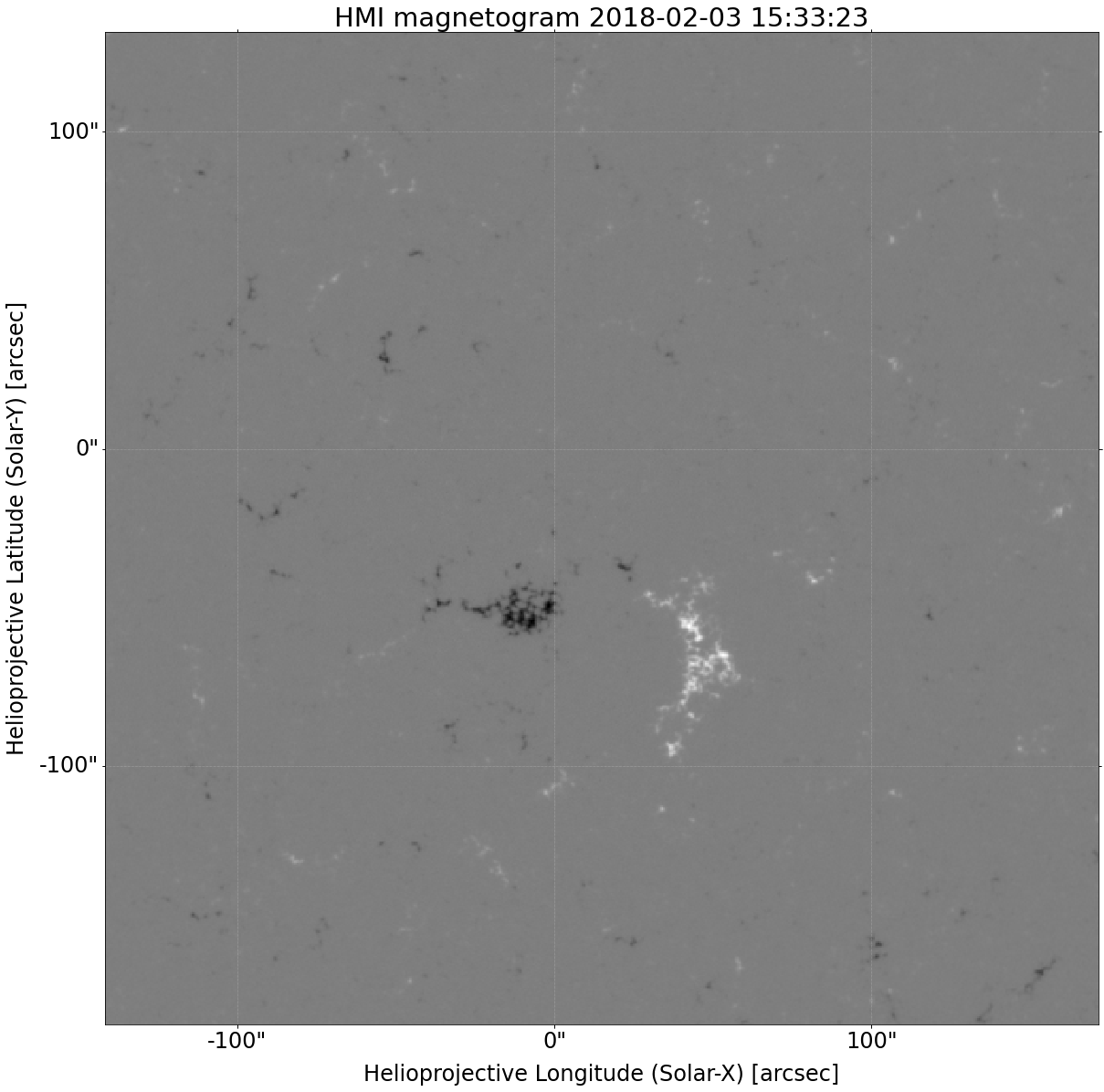}\end{center}
		\caption{Starting magnetogram used for the prototype test case.}\label{fig:magnetogram}
	\end{figure}

	Where $j$ indexes the loops and $i$ indexes each pixel location. Normally, two indexes such as $i_x$ and $i_y$ are used to represent the pixel locations, but here $i$ represents the flattened pixel locations in the usual way, i.e., $i = i_xn_y+i_y$. This is a one-to-one mapping between the single index $i$ and the two indices $i_x, i_y$. Therefore, $E_i$ is entirely equivalent to the more conventional notation $E(i_x,i_y)$ for the intensities at each pixel location in the image; likewise, $P_{ij}$ is equivalent to $P_j(i_x,i_y)$, so that
	
	\begin{equation}
		E(i_x,i_y) = \sum_j P_j(i_x,i_y) c_j
	\end{equation}

	\noindent is equivalent to Equation \ref{eq:linear_loop_response}. The question then is, what is the $P_{ij}$ matrix, and how is it obtained? It is the response of each pixel (indexed by $i \equiv i_x n_y + i_y$) to each loop (indexed by $j$). There are a number of ways to arrive at this matrix; the following is a prototype implementation, which has its origins in the SynthesizAR framework \citep{BarnesEtal_ApJ2016I,BarnesEtal_ApJ2016II}. It begins as follows:

	\begin{enumerate}
		\item Perform a potential field extrapolation from a longitudinal AIA magnetogram (see in Figure \ref{fig:magnetogram}).
		\item Trace a set of field lines and lay out a three-dimensional grid of voxels (just as with the pixels, the voxels are addressed with a single flattened index, $k$).
		\item Identify each voxel with the field line it is closest to. From this identification, the connection between the loops and pixels on the image plane becomes straightforward (see below).
		\item Compute the (longitudinal) arc length of each voxel with respect to its field line.
	\end{enumerate}

	Much of this implementation grew out of a project studying coronal heating, where it is being used to compute three-dimensional coronal loop heating distributions. The implementation and application will be described in more detail in an upcoming publication on that project. We now have loop index ($j$), arc length ($l_k$), and coordinates including height ($h_k$, direct from the voxel grid) of each voxel. This is illustrated in Figure \ref{fig:xyproject}. All of these can be used to compute emission profiles for each voxel and loop. To relate these coordinates and loop indices to the physics of the loop and the emission of the voxels, we can, for example, compute a voxel intensity/emissivity, $e_{kj}$, given a loop temperature profile $T_j(l_k)$ that is a function of length and a pressure profile $P_j(h_k)$ that is a function of height, and a temperature response function $R(T)$:

	\begin{figure}[!htbp]
		\begin{center}\includegraphics[width=0.33\textwidth]{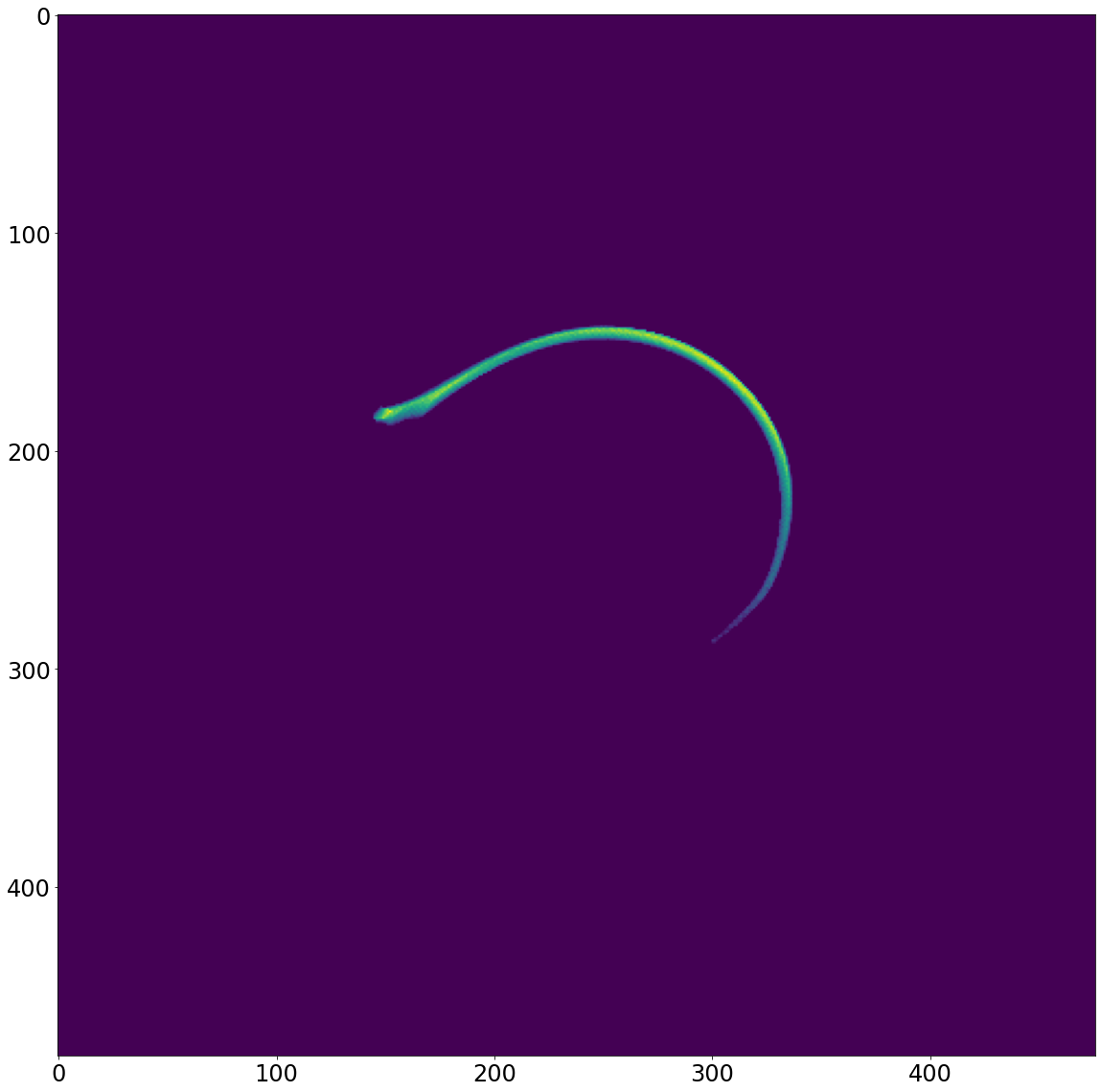}\includegraphics[width=0.33\textwidth]{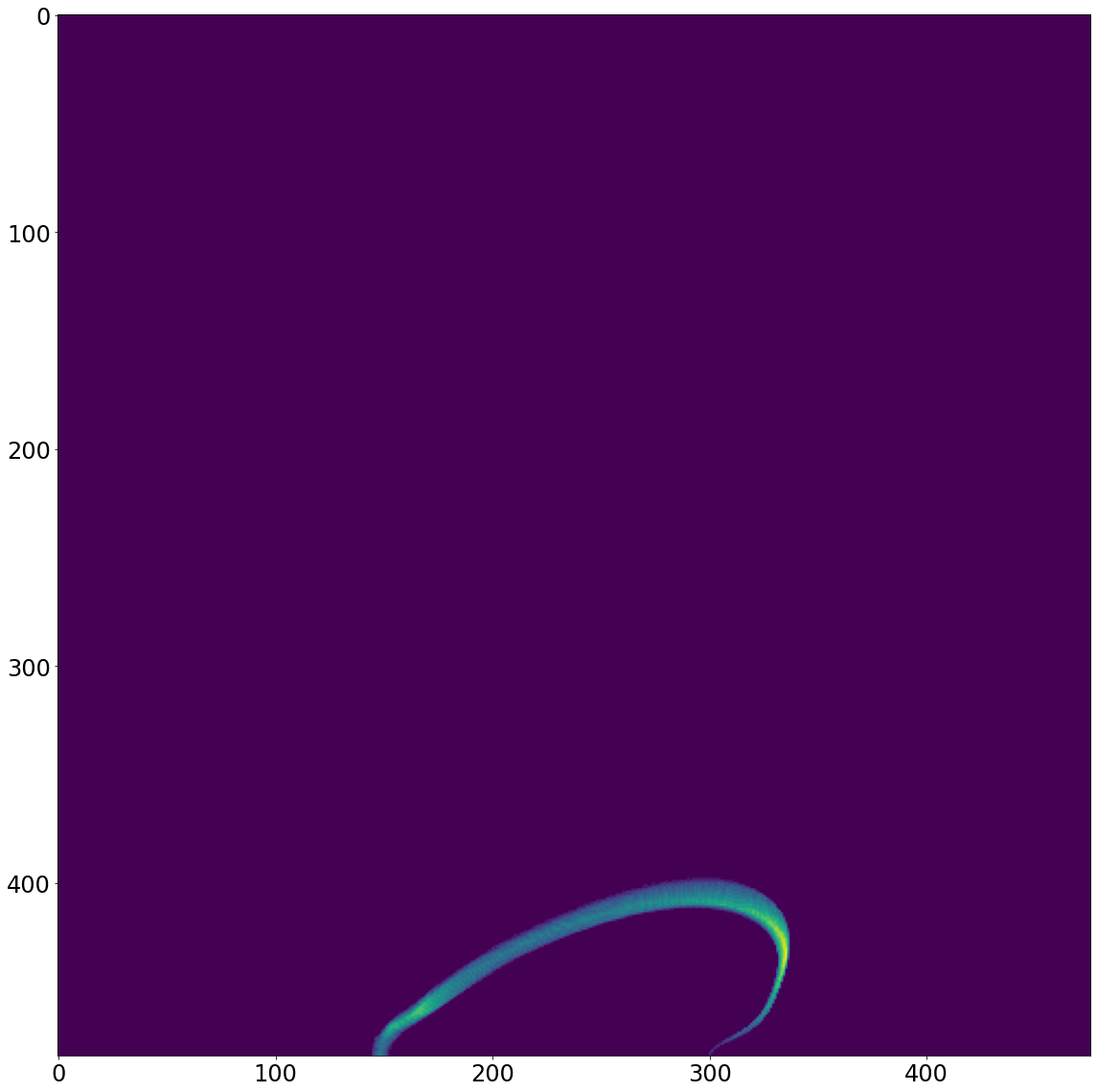}\includegraphics[width=0.33\textwidth]{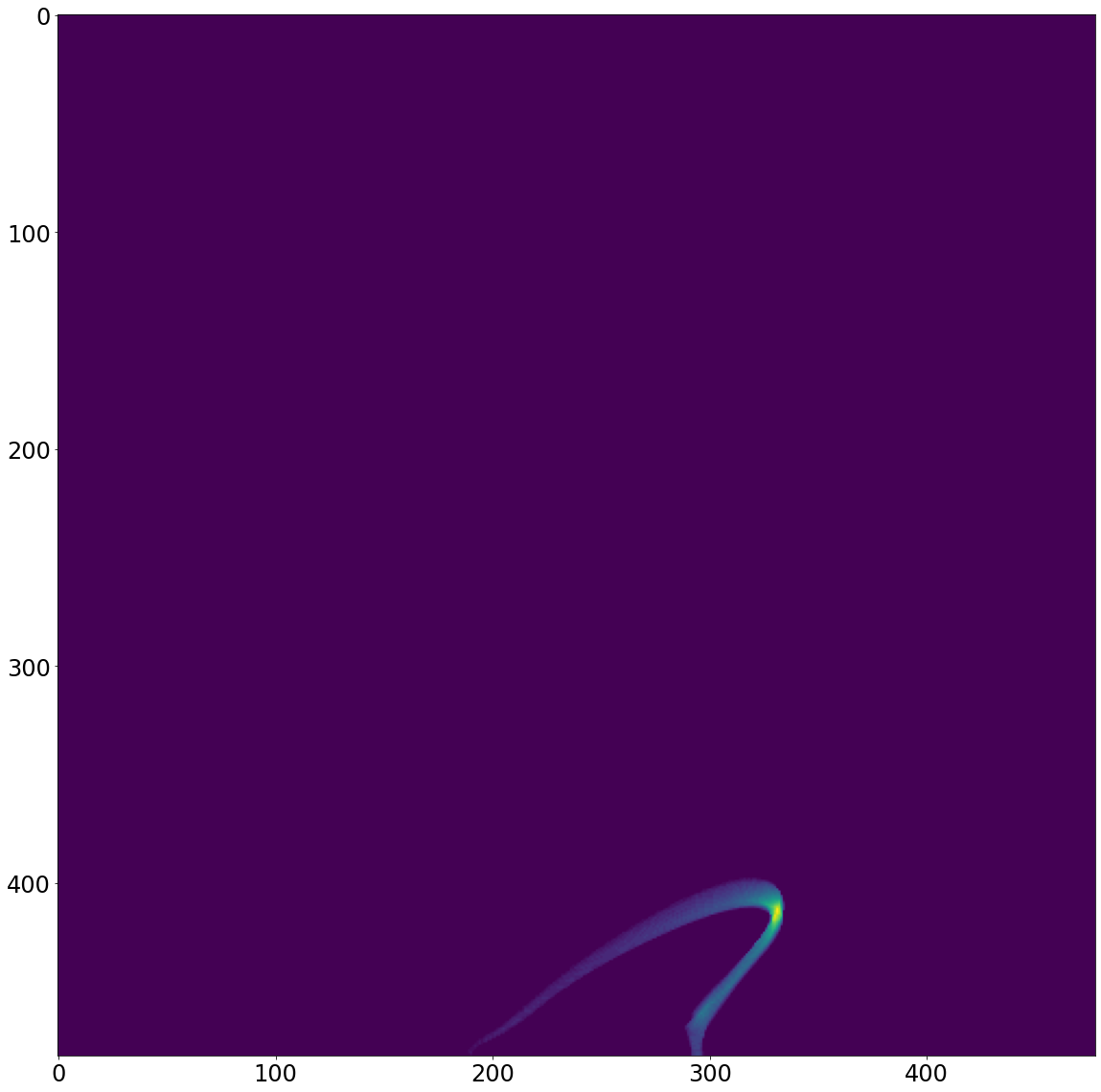}\end{center}
		\caption{Voxels for an example loop, viewed from above (left) and from the x-z and y-z planes (center, right).}\label{fig:xyproject}
	\end{figure}

	\begin{equation}
		e_{kj} = c_0R(T_j(l_k))\Big(\frac{P_j(h_k)}{T_j(l_k)}\Big)^2,
	\end{equation}

	\noindent where we have used the ideal gas law to replace density with pressure and temperature, and $c_0$ is a proportionality constant. Now, we make the simplification that all loops have the same temperature profile except with a different scaling coefficient, $c'_j$,

	\begin{equation}
		T_j(l_k) = c'_jt_p(l_k),
	\end{equation}

	\noindent The pressures assuming constant gravity can likewise be approximated by
	\begin{equation}
		P_j(h_k) = p_j \exp\Big[-\frac{h_k}{c_h T}\Big],
	\end{equation}	
	Where $c_h T$ is a scale height and $p_j$ is the pressure at the base of the loop (for sufficiently hot or low loops, this reduces to uniform pressure), so that the voxel intensities are given by 

	\begin{equation}\label{eq:voxel_intensities}
		e_{kj} = c_0R(c'_j t_p(l_k))\Big(\frac{p_j \exp\big[-h_k/(c_h T)\big]}{c'_j t_p(l_k)}\Big)^2.
	\end{equation}

	The relationship between these voxel intensities and the pixel intensities, $E_i$, is essentially just the pixel PSF and apodization function. This can be applied by a linear convolution, or equivalently, a matrix multiplication. Moreover, because the corona is optically thin, the radiative transfer can be approximated by a simple projection operator. Therefore, the relationship between voxel intensities and pixel intensities can can be represented by a `spatial response matrix', $S_{ik}$:

	\begin{equation}\label{eq:forwardproblem_full}
		E_i = \sum_{jk} S_{ik} e_{kj} = \sum_{jk} S_{ik} c'_0R(c'_j t_p(l_k))\Big(\frac{p_j \exp\big[-h_k/(c_h T)\big]}{c'_j t_p(l_k)}\Big)^2.
	\end{equation}

	This response matrix is straightforward to compute -- it is the combination of the image plane projection, instrument PSF, and pixel apodization functions -- so it will not be detailed further here. Equation \ref{eq:forwardproblem_full} is not quite the linear form that we seek (Equation \ref{eq:linear_loop_response}) -- one coefficient, $c_j'$, is buried in the response function, and it appears as its reciprocal squared elsewhere. These we can fix if the temperature response function is of the form 

	\begin{equation}\label{eq:powerlaw_response}
		R(T) = C_T T^{\mu} = C_T (c'_j)^\mu t_p(l_k)^\mu,
	\end{equation}

	\noindent where $C_T$ is a proportionality constant. This `power law' form is also similar to the response functions of X-ray passbands such as observed by Hinode XRT. Alternatively, Differential Emission Measures \citep[DEMs; e.g.,][]{Cheungetal15,PlowmanCaspi2020} allow this (or any other response function) to be obtained from the AIA data, provided the temperatures being observed are within the range of AIA's temperature response functions. This is the method employed by the paper's prototype: obtain an optically thin `synthetic channel' image with this response function by first computing a DEM from the AIA data and then integrating the DEM against the above R(T) -- The top left panel of Figure \ref{fig:prototype_results} shows a synthetic channel image with $\mu=2$  (This synthetic channel is the column integral of the pressure squared, just as the total emission measure is the column integral of the density squared, which may be interesting in its own right). This power law form is an important innovation, because it allows us to essentially linearize the parameter search space, as we now show. Equation \label{eq:forwardproblem_full} becomes
	\begin{equation}\label{eq:fullform_singleterm}
		E_i = \sum_{jk} S_{ik} e_{kj} = \sum_{jk} S_{ik} c_0C_T (c'_j)^{\mu-2} [t_p(l_k)]^{\mu-2} [p_j \exp\big[-h_k/(c_h T)\big]]^2,
	\end{equation}

	\noindent which allows us to match Equation \ref{eq:linear_loop_response} by identifying

	\begin{equation}\label{eq:overall_coefficient}
		c_j = p_j^2(c'_j)^{\mu-2}
	\end{equation}
	\noindent and
	\begin{equation}
		P_{ij} = \sum_k c_0C_T [t_p(l_k)]^{\mu-2} \exp\big[-2h_k/(c_h T)\big].
	\end{equation}
	Each row of $P_{ij}$ is an image of the $j$th loop as seen by the instrument. In fact, in the case where $\mu=2$ and the scale height is large (uniform pressure), these images are essentially identical to those shown in Figure \ref{fig:xyproject}: $\mu=2$ cancels out the temperature profile dependence entirely, leaving a reconstruction that is exclusively a function of pressure.

	\section{Discussion and refinements}

	A crucial innovation in this approach is that the relationship between the observables (the $E_i$) and the unknowns ($c_j$) is linear. With the equations in this form, the best fit loop coefficients, $c_j$, can be determined by linear least squares $\chi^2$ minimization. This is the core of the solution to the 3D plasma reconstruction problem; the following innovations are added to complete the solution:

	\begin{itemize}
		\item Determining the thermodynamic state of each voxel (e.g., its temperature and density) involves at least two unknowns (e.g., temperature and density), and therefore two equations per voxel. Equation \ref{eq:voxel_intensities} provides just one for a given input image/response function. To provide an additional equation, the reconstruction is performed for two distinct images (and corresponding response functions) independently, which yields two equations for each voxel. For example, performing a reconstruction with a $\mu=2$ image provides the pressure at each voxel, while performing it for a $\mu=3$ image provides the product of the pressure and the temperature at each voxel; the ratio of these two therefore provides the temperature at each voxel.
		\item Assuming a single temperature profile for all loops may be limiting; However, a superposition of a second or third profile with its own independent coefficient for each loop can be added. Allowing a larger number of such profiles enables a fully general reconstruction to be attempted, but at the expense of ill-posedness. See additional discussion below.
		\item The pressure profile has a temperature dependance (or, equivalently, a scale height must be assumed). This can also be mitigated with the use of a second profile and coefficient as discussed above, or alternatively with an iterative approach; e.g., begin with an initial guess for the scale height in all loops, carry out the reconstruction, use the results to update the guess each loop's scale height, and repeat. The prototype demonstration shown below uses the first approach.
	\end{itemize}
	
	This approach resolves the ill-posedness which has previously thwarted three-dimensional reconstruction techniques. Rather than having $n_x\times n_y \times n_z$ unknowns to recover from a two-dimensional image with $n_x \times n_y$ pieces of information, there are only have of order $n_\mathrm{loops}$ pieces of information to recover. This is of order $n_x \times n_y$ or fewer, since almost all loops are rooted in the photosphere. The preliminary reconstructions shown next have used of order 2,000 loops, which is already adequate to resolve the structure in the image. To further constrain the solution, positivity of the $c_j$ is enforced using the same technique demonstrated in \cite{PlowmanCaspi2020}. Even with multiple components per loop, and varying $\alpha$ parameters in a force-free magnetic field solution (discussed below), there will still be only have of order $10^4$ unknowns, as opposed to of order $10^6$ pieces of information per image and several images (one magnetogram and at least two optically thin images). The problem can therefore remain well-constrained even with mutiple profiles and a nonlinear force-free field (discussed next).

	An additional means of leverage which can be employed are ill-posedness are regularization-type constraints. These include general mathematical constraints such as the $l^1$ and $l^2$ norms and smoothness \citep[e.g.,][]{Cheungetal15, PlowmanCaspi2020}. The prototype has not had to rely on these, except in as much as they are implicit in the \texttt{bicgstab} sparse solver's tolerance level (we also normalize the $P_{ij}$ so that the entries of each loop sum to one). They can therefore be kept `in our back pocket' in case of difficulty with ill-posedness. Moreover, the fact that the methodology laid out in this paper can obtain the full 3D temperature, density, and magnetic field will also enable an even more powerful class of constraints, if need be: The physical equations governing the plasma. For example, penalty functions can be applied for any deviation of physically conserved quantities in a time-varying solution, or (if applicable) the MHD equations.


	\subsection{Multiple Profiles}

	The dependence on a single temperature profile can be avoided by providing multiple profiles for each loop. Essentially, rather than the loop (and its voxels) being defined by a single temperature profile, density profile, and response function, as in Equation \ref{eq:voxel_intensities}, we use a superposition of them:
	
	\begin{equation}\label{eq:fullgeneral_loopprofile0}
		e_{kj} = \sum_m c^m_j f_m(T(l_k),l_k,h_k,\dots),
	\end{equation}
	Where each $f_m$ is similar to the right-hand side of Equation \ref{eq:voxel_intensities}. These can be functions of any known geometrical properties of the loop and voxels (length, height, area, etc), but the same set must apply for all loops. With enough coefficients and functions, this can in principal be used to represent any configuration of the plasma within the volume, but there is not enough information in the images to allow a large number of coefficients per loop to be recovered (just as in the general $n_x\times n_y \times n_z$ problem).
	
	In moderation, though, the multiple profile approach can remain physically valid while not introducing ill-posedness. Consider applying Equation \ref{eq:fullform_singleterm} with $\mu=2$ but with two pressure scale heights rather than one. Folding all of the constants into the coefficient, the result is a simple two-term expression:
	
	\begin{equation}\label{eq:2comp_profile}
		E_i = \sum_{jk} S_{ik} \Big(c_j^0 \exp\big[-2h_k/(c_h T_0)\big] +  c_j^1 \exp\big[-2h_k/(c_h T_1)\big]\Big),
	\end{equation}
	
	where $c_j^0$ is the coefficient for the profile with scale height $c_h T_0$ and $c_j^1$ is the coefficient with scale height $c_h T_1$. The reconstruction can then interpolate between the two scale heights as necessary. The first example shown in Section \ref{sec:results} uses these two components.

	In order to link $e_{kj}$ to the emission physics of the problem, the $f_m$ functions and the temperature response function $R(T)$, must be chosen such that

	\begin{equation}\label{eq:fullgeneral_loopprofile}
		R(T(l))n^2(l) \approx c_j^1 f_1(l,h(l),a(l)) + c_j^2 f_2(l,h(l),a(l)) + \dots.
	\end{equation}
	The coefficients $c_j^1,c_j^2,$ etc can vary from one loop to the next, but the same functions $f_1, f_2,$ etc must apply for all loops -- irrespective of their heating amount or frequency. They can be any positive-definite functions, but they must apply for all loops. The temperature response function $R(T)$ may likewise be almost any positive-definite function (thanks to DEMs and the broad temperature coverage of instruments such as AIA), which is a great advantage: In equations \ref{eq:powerlaw_response} and \ref{eq:overall_coefficient}, it allowed us to essentially linearize the parameter search space.

	Equations \ref{eq:fullgeneral_loopprofile0} and \ref{eq:fullgeneral_loopprofile} are the most general form of the problem; Section \ref{sec:formalism} has already demonstrated how to these functions can be chosen in a way that is physically motivated and satisfies Equation \ref{eq:fullgeneral_loopprofile}. For example, Equation \ref{eq:2comp_profile} corresponds to $R(T) = T^2$, $f_1 = \exp\big[-2h_k/(c_h T_0)\big]$, and $f_2 = \exp\big[-2h_k/(c_h T_1)\big]$. That is the $\mu=2$ case where the temperature profile dependence vanishes (assuming ideal gas law). Equation \ref{eq:fullform_singleterm}, on the other hand, corresponds to $R(T) = T^\mu$, and $f_1 = t_p(l)^{\mu-2}\exp\big[-2h_k/(c_h T_0)\big]$, where $t_p$ is a temperature profile. This temperature profile and the overall profile $f_1$ are somewhat interchangeable in this case. 
	
	\section{Preliminary Results, Implications \& Next Steps}\label{sec:results}

\begin{figure}[!htbp]
		\begin{center}\includegraphics[width=\textwidth]{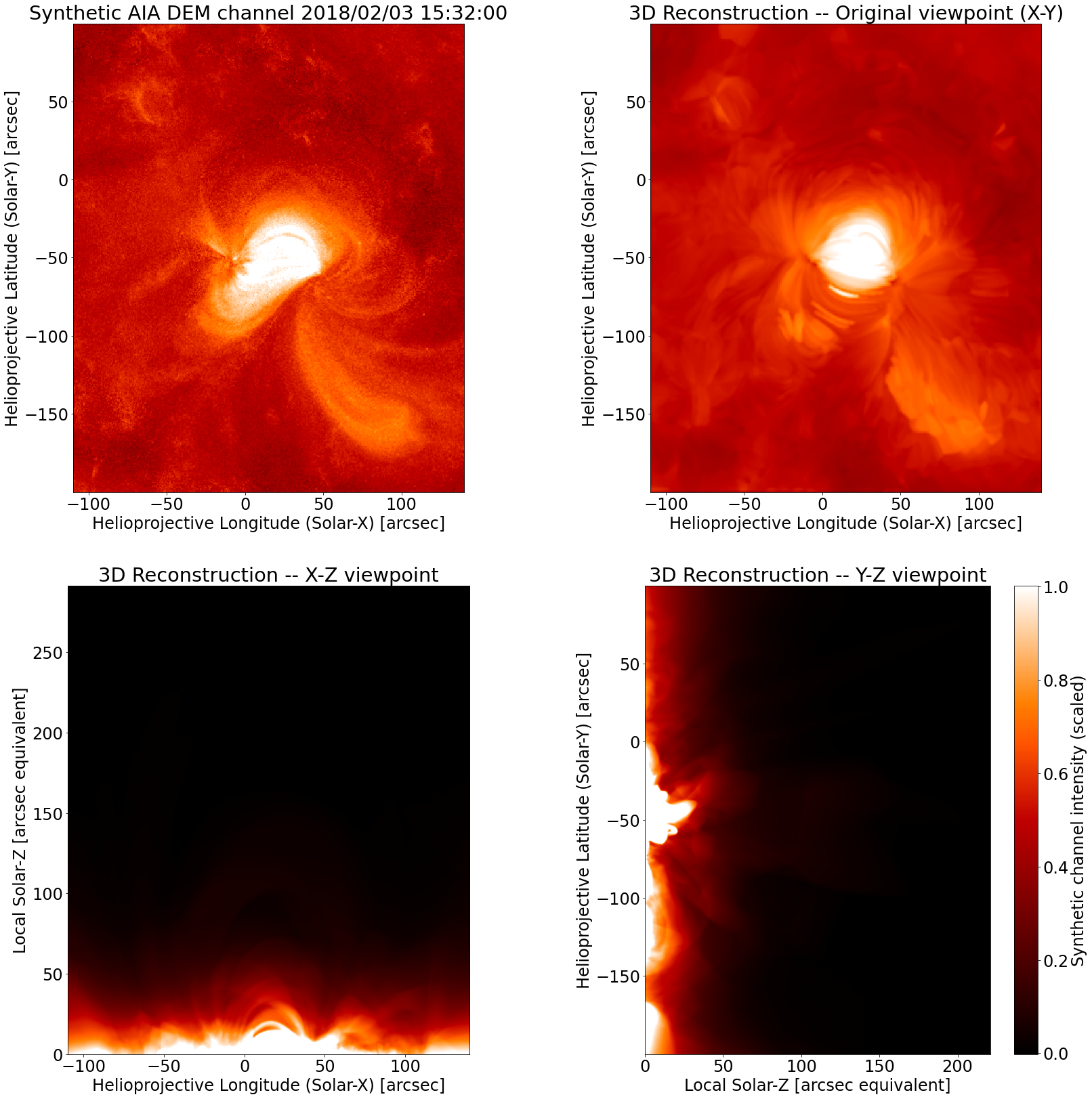}\end{center}
		\caption{Top left shows AIA based image (Computed from DEM with $\mu=2$ in R(T)) for the test reconstruction, corresponding to the magnetogram shown in Figure \ref{fig:magnetogram}. Other panels show the reconstruction: top right for the same perspective as the AIA image, while bottom panels show the x-z and y-z planes (left and right, respectively). The reconstruction is computed exclusively from SDO data and its viewpoint.}\label{fig:prototype_results}
	\end{figure}

	At this point, we have demonstrated how to determine the plasma temperature/pressure/density which give the best fit $\chi^2$ assuming a magnetic magnetic field skeleton. Results of a prototype implementation of this methodology are shown in Figure \ref{fig:prototype_results}. This preliminary implementation was done in Python, with the actual inversion and solution of the linear least squares problem done using the \texttt{bicgstab} sparse matrix solver algorithm as implemented by the \texttt{scipy.sparse.linalg} package. There are 1500 loops in this solution with two profiles per loop, so the size of the matrix to be inverted is $3000 \times 3000$. One profile has a scale height of 30 Mm, while the other has a scale height of 150 Mm. The \texttt{bicgstab} algorithm is very fast for this matrix size, returning an inversion in a fraction of a second on a 15 Watt laptop CPU. In other tests with up to $\sim 20000$ loops, its performance has remained good. In the current prototype most of the computational time is occupied by computing the mappings from loops to voxels and from voxels to pixels, rather than by the actual inversion; however, there is ample room for optimization in those initial steps.

	The performance of this reconstruction illustrates the potential of this algorithm, both to recover the 3D plasma structure and to constrain the magnetic field. In the top half of Figure \ref{fig:prototype_results} (above latitude -50"), the field appears closer to potential, and the reconstruction is quite good. In the lower half, the field appears non-potential, and the reconstruction struggles -- better field extrapolations in that region produce better reconstructions, so the quality of the plasma reconstruction (i.e., the $\chi^2$ between the reconstruction and the optically thin data) can be used to refine the field extrapolation.

\begin{figure}[!htbp]
		\begin{center}\includegraphics[width=\textwidth]{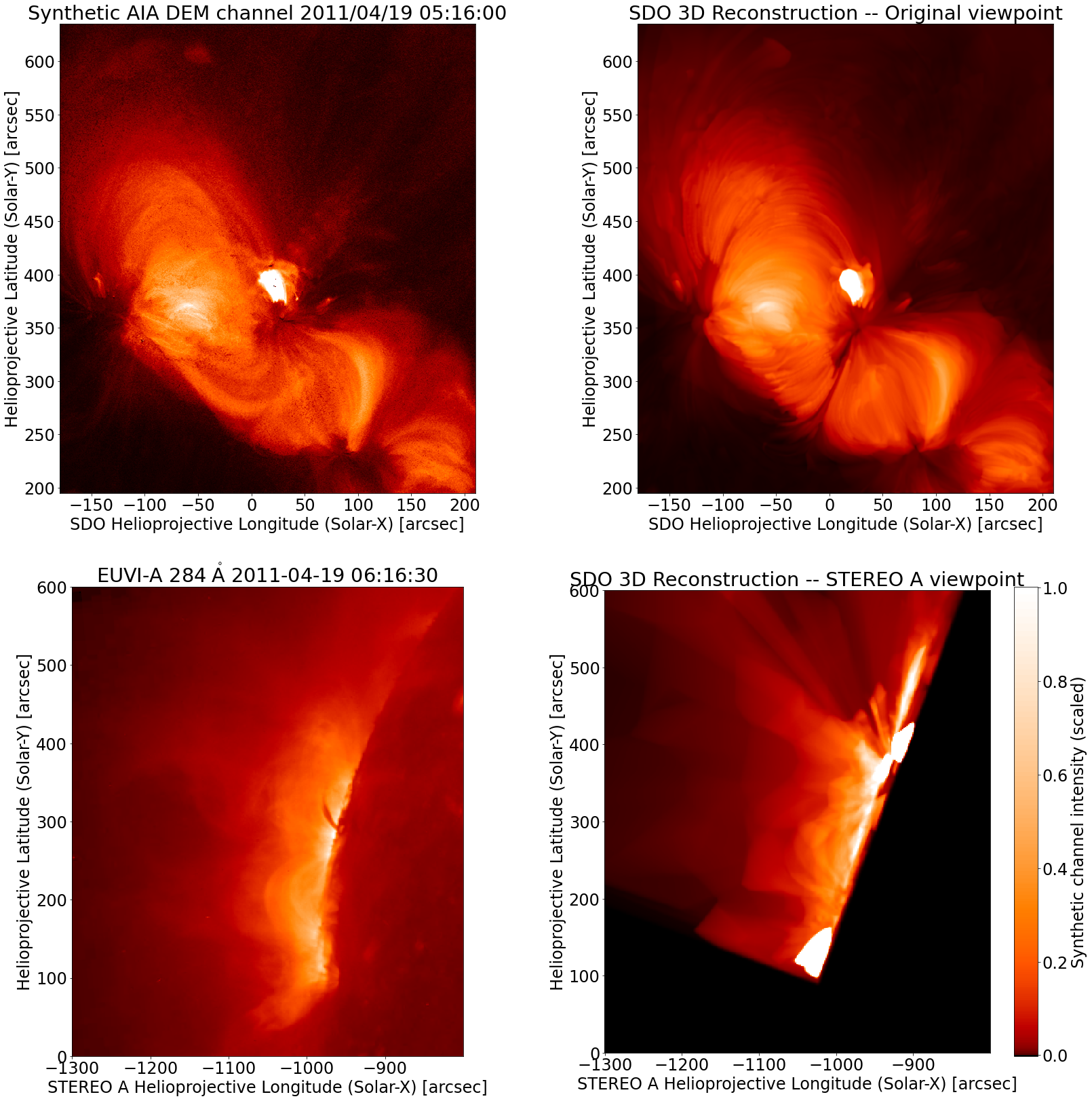}\end{center}
		\caption{Top left shows AIA based image (Computed from DEM with $\mu=2$ in R(T)) for the test reconstruction. Top right shows the reconstruction from the original (SDO) perspective. Bottom left shows the corresponding STEREO 284 \AA\ observations. Bottom right shows the reconstruction from STEREO's perspective: The reconstruction is computed exclusively from SDO data and its viewpoint, yet many features in the reconstructed image clearly match those in the STEREO data.}\label{fig:prototype_results_STEREO}
	\end{figure}

	To investigate the quality of the reconstruction's alternative perspectives, another reconstruction is performed for a second, much larger region, which was viewed by the STEREO spacecraft when they were $90^\circ$ from the Earth-Sun line (i.e., orthogonal). 2000 loops were used, and as in the previous case, the reconstruction was done with two profiles per loop. However, rather than having the profiles represent two scale heights, we use the multiple profiles to allow us to model asymmetric loops: one profile represents the `left' side of the loop, while the other represents the `right', with the profiles blended equally in the middle. A single scale height of 120 Mm was used for both profiles. The results are shown, in comparison with STEREO A 284 \AA\ , in Figure \ref{fig:prototype_results_STEREO}; the reconstruction itself is done exclusively with SDO data and its perspective. As before, areas which appear close to potential show a clear correspondence between both the STEREO and the AIA observations. This is primarily in the lower right part of the AIA image, and to a lesser extent in the active region core, in this case. This is impressive considering the active region is complex, dynamic, and large enough that the solar curvature is significant (the prototype assumes a planar surface geometry).
	
	This demonstrates the capability to use a magnetic field extrapolation and optically thin images to determine a three-dimensional plasma or emission measure distribution. As part of this process, simulated images (i.e., the $E_i$ of Equation \ref{eq:linear_loop_response} are produced which are compared directly with the input optically thin image via $\chi^2$. The reconstruction produces the best fit emission and plasma distribution given the input magnetic field. In the case shown here, the reconstructed images clearly resemble the original, but there remains a significant mismatch which is clearly due to differences in magnetic field structure. Rather than being a problem, this represents a best of both worlds scenario: The reconstruction of the optically thin emission is robust enough to find a solution which resembles the original even in the absence of magnetic field differences, {\em and} there remains substantial additional information in the residuals which can be used to constrain the three-dimensional coronal magnetic field.
	The next steps in this research will investigate this coronal magnetic field optimization and constraint problem; The following concluding summary describes that can be accomplished.
	
	To apply these constraints to the field, a parameterized non-potential field extrapolation will be required. The logical choice will be first linear, and then nonlinear, force-free field extrapolations \citep[LFFF/nLFFF; e.g., ][]{WiegelmannSakurai_LRSP2012}. The linear force-free field is the simplest possible case to model, having a single tunable $\alpha$ parameter. This makes for a simple search: simply try various values of the alpha parameter, and perform the 3D reconstruction performed in this paper, then refine and repeat based on the residual $\chi^2$ with the emission images.  For each of these values we will carry out the 3D emission reconstruction process described in this paper, computing the residual $\chi^2$ of the emission for each. Beginning with the $\alpha$ that produces the best of these, we will then be used to refine the $\alpha$ parameter, further minimizing $\chi^2$.  
	
	NLFFFs are an extension of LFFFs with $\alpha$ allowed to vary over space. To search for nLFFFs, the best-fit LFFF (found as described above) can be used as a starting guess. Spatially varying pertubations to $\alpha$ can then be applied using (for example) a harmonic expansion, and then the coefficients of the perturbation expansion can be searched for the best-fit to the emission data (again, using the 3D reconstruction as part of the search) by, for example, a Levenberg-Marquardt algorithm \citep{nr}. The performance of the 3D reconstruction makes its use feasible in this way, inside another optimization algorithm. 

	The methodology laid out in this paper represents a tractable way forward to solve one of the most difficult and enduring data analysis problems in solar physics. The prototype demonstration shown here already produces results which clearly resemble the real sun, and it is done without appreciable regularization and with only a potential magnetic field structure. It is likely that regularization and the correct field structure will allow it to produce highly usable results, even with just one perspective. With multiple persepectives \citep[e.g., from Solar Orbiter][]{Orbiter_ForveilleShore2020}, this is almost guaranteed. These high quality three-dimensional plasma and magnetic field reconstructions would have impact across many disciplines in solar physics, and greatly advance the science objectives of missions such as Orbiter, SDO, and {\em Hinode} -- from identifying the locations of coronal heating, to understanding the connections from the solar surface to the heliosphere, to predicting space weather.
		
	\acknowledgements{This work was funded by NASA grant 80NSSC17K0598. The author thanks Will Barnes, whose SyntheizAR framework provided an important starting point for the implementation, as well as Doug Nychka and Natasha Flyer, who provided initial suggestions on the inversion formalism and computational tools. The implementation makes extensive use of SunPy.}

\bibliographystyle{aasjournal}
\bibliography{reconstruction_paper0}

\end{document}